\documentclass[pra,aps,twocolumn,epsfig,superscriptaddress,showpacs]{revtex4}

\usepackage{mathrsfs}
\usepackage{graphicx}
\usepackage{amsfonts}
\usepackage{amsmath}
\usepackage{amssymb}

\def\endproof{\vrule height6pt width6pt depth0pt}

\begin{document}


\title{Quantum contextuality for a relativistic spin-1/2 particle}


\author{Jing-Ling Chen}
 \email{chenjl@nankai.edu.cn}
 \affiliation{Theoretical Physics Division, Chern Institute of Mathematics, Nankai University,
 Tianjin 300071, People's Republic of China}
 \affiliation{Centre for Quantum Technologies, National University of Singapore,
 3 Science Drive 2, Singapore 117543}

\author{Hong-Yi Su}
 \affiliation{Theoretical Physics Division, Chern Institute of Mathematics, Nankai University,
 Tianjin 300071, People's Republic of China}
 \affiliation{Centre for Quantum Technologies, National University of Singapore,
 3 Science Drive 2, Singapore 117543}

\author{Chunfeng Wu}
 \affiliation{Centre for Quantum Technologies, National University of Singapore,
 3 Science Drive 2, Singapore 117543}

\author{Dong-Ling Deng}
 \affiliation{Department of Physics and MCTP, University of Michigan,
 Ann Arbor, Michigan 48109, USA}

\author{Ad\'{a}n Cabello}
 \email{adan@us.es}
 \affiliation{Departamento de F\'{\i}sica Aplicada II, Universidad de
 Sevilla, E-41012 Sevilla, Spain}
 \affiliation{Department of Physics, Stockholm University, S-10691
 Stockholm, Sweden}

\author{L. C. Kwek}
 \email{kwekleongchuan@nus.edu.sg}
 \affiliation{Centre for Quantum Technologies, National University of Singapore,
 3 Science Drive 2, Singapore 117543}
 \affiliation{National Institute of Education and Institute of Advanced Studies,
 Nanyang Technological University, 1 Nanyang Walk, Singapore 637616}

\author{C. H. Oh}
 \email{phyohch@nus.edu.sg}
\affiliation{Centre for Quantum Technologies, National University of
Singapore, 3 Science Drive 2, Singapore 117543}
\affiliation{Department of Physics, National University of
Singapore, 2 Science Drive 3, Singapore 117542}

\date{\today}


\begin{abstract}
The quantum predictions for a single nonrelativistic spin-1/2
particle can be reproduced by noncontextual hidden variables. Here
we show that quantum contextuality for a relativistic electron
moving in a Coulomb potential naturally emerges if relativistic
effects are taken into account. The contextuality can be identified
through the violation of noncontextuality inequalities. We also discuss
quantum contextuality for the free Dirac electron as well as the
relativistic Dirac oscillator.
\end{abstract}


\pacs{03.65.Ta, 03.65.Ud, 03.30.+p}

\maketitle


\section{Introduction}


Noncontextual hidden variable theories assume that the results of
measurements are independent of which other compatible observables
are jointly measured~\cite{Mermin}. The Kochen-Specker (KS) theorem
\cite{KS67} states that noncontextual hidden variables cannot
reproduce the predictions of quantum mechanics for systems of
dimension $d \ge 3$. This is known as quantum contextuality and is
state-independent: For any dimension $d \ge 3$, there are universal
sets of quantum observables which prove contextuality for any state
of the system. Moreover, it has recently been shown that for any
physical system of $d \ge 3$ there is an inequality satisfied by any
noncontextual hidden variable theory but which is violated, for any
quantum state, by a universal set of quantum observables
\cite{Cabello08,BBCP09}. Recent experiments have confirmed
state-independent quantum contextuality \cite{KZGKGCBR09,ARBC09}.
The significance of these results can be summarized in the statement
that, for systems of dimension higher than two, there are no
``classical'' (i.e., noncontextual) states \cite{BBCP09}.

However, there still remains a debate on whether quantum
contextuality can be defined on systems of dimension two such as a
single spin-1/2 particle. Using the standard approach of the KS
theorem, based on von Neumann projective measurements, it is
impossible to define contextuality on a single qubit, since every
qubit observable is only compatible with itself and hence only
appears in one measurement context. By adopting positive
operator-valued measurements, Cabello~\cite{Cabello03} and
Nakamura~\cite{Nakamura03} have shown that a single qubit exhibits a
form of contextuality. However, Grudka and Kurzy\'nski \cite{GK08}
have criticized this approach by pointing out that the contextuality
in Ref.~\cite{Cabello03,Nakamura03} is different than the contextuality
in the KS theorem. The issue of whether a single spin-1/2 particle
can exhibit KS contextuality remains a pending problem.

Here we adopt a completely different perspective. We start with a specific physical qubit: the spin of an electron. Within the framework of nonrelativistic quantum mechanics, the spin of an electron is treated as a two-dimensional system and does not exhibit KS contextuality. However, the situation dramatically changes when special relativity is taken into account.

By requiring the relativistic wave-equation to be a first-order
differential equation with respect to time and spatial coordinates
and Lorentz-invariant under space-time transformations, Dirac
discovered his famous equation. For an electron moving in a
potential $V(r)$, its relativistic Hamiltonian is given by
\begin{equation}
 \label{HRHA1}
 H=c\ \vec{\alpha}\cdot \vec{p}+\beta Mc^2+V(r),
\end{equation}
with $\vec{\alpha} =\sigma_x\otimes\vec{\sigma}$ and $\beta=
\sigma_z\otimes\openone$, $\vec{p}$ being the linear momentum,
$\vec{r}$ the coordinate, $r=|\vec{r}|$, $M$ the rest mass of the
electron, $c$ the speed of light in vacuum, $\vec{\sigma}$ the
vector of Pauli matrices, and $\openone$ the $2 \times 2$ identity
matrix. The angular momentum should be a conserved quantity for the
Hamiltonian $H$. However, the orbital angular momentum
$\vec{L}=\vec{r}\times \vec{p}$ does not commute with $H$ unless one
adds it up with a quantity $\vec{S}=\frac{\hbar}{2}\vec{\Sigma}$,
with $\vec{\Sigma}=\openone\otimes\vec{\sigma}$ and $\hbar$ the
Planck constant. The quantity $\vec{S}$ is nothing but the intrinsic
spin angular momentum. From $\vec{S}^2=\frac{3}{4}\hbar^2$ one may
determine that its spin value is $\frac{1}{2}$. Consequently, the
spin-1/2 angular momentum has a natural origin within relativistic
quantum mechanics. According to Landau and Lifshitz, ``this property
of elementary particles [the spin] is peculiar to quantum mechanics
(\ldots) and therefore has in principle no classical
interpretation''~\cite{Landau}. An immediate question arises: Is
there KS contextuality for a single spin-1/2 particle moving in the
potential $V(r)$ within the framework of relativistic quantum
mechanics?

In this work, we provide an affirmative answer to this question and
demonstrate that contextuality of a single hydrogen atom (i.e., a
relativistic electron moving in the Coulomb potential) naturally
emerges from a relativistic treatment. We prove that all eigenstates
of the relativistic hydrogen atom violate a noncontextuality
inequality. The contextuality of the free Dirac electron and the
Dirac oscillator is also discussed based on the measurability of the
observables.


\section{Quantum contextuality for the relativistic hydrogen atom}


It is interesting to study the quantum contextuality of a
relativistic electron moving in a Coulomb potential. This is just a
model of a single relativistic hydrogen atom (RHA). The
corresponding Dirac Hamiltonian reads
\begin{eqnarray}
 H_{\rm rha}=c\ \vec{\alpha}\cdot \vec{p}+\beta Mc^2-\frac{\hbar c a}{r},
\end{eqnarray}
with $a=e^2/\hbar c\simeq1/137.036$ being the fine structure
constant, and $e$ the electric charge. The energy spectrum is given
by the Sommerfeld formula
\begin{eqnarray}
 \label{E}
&&\frac{E}{Mc^2}=\left(1+\frac{a^2}{(n-|\kappa|+\sqrt{\kappa^2-a^2})^2}\right)^{-1/2},
\\
&&|\kappa|=(j+1/2)=1,2,3,\ldots,\quad n=1,2,3,\ldots\nonumber
\end{eqnarray}
The common eigenfunctions of $\{ H_{\rm rha}, \vec{J}^2, J_z\}$ are
twofold Krammer's degeneracies \cite{Bethe,AJP}, i.e.,
\begin{subequations}
 \label{E1}
 \begin{align}
 &|\psi^+_{njm_j}(\vec{r})\rangle=\frac{1}{\sqrt{\cal
N}}\left(\begin{matrix}
if(r)\phi^A_{jm_j}\\
g(r)\phi^B_{jm_j}
\end{matrix}\right),\\
&|\psi^-_{njm_j}(\vec{r})\rangle=\frac{1}{\sqrt{\cal
N}}\left(\begin{matrix}
if(r)\phi^B_{jm_j}\\
g(r)\phi^A_{jm_j}
\end{matrix}\right),
\end{align}
\end{subequations}
where $K |\psi^\pm_{njm_j}(\vec{r})\rangle= \pm |\kappa|
|\psi^\pm_{njm_j}(\vec{r})\rangle$, with
$K=\beta(\vec{\Sigma}\cdot\vec{L}/\hbar+1)$ being the Dirac
operator, $K^2=\vec{J}^2/\hbar^2+1/4$, and $\vec{J}=\vec{L}+\vec{S}$
the total angular momentum operator.
$|\psi^\pm_{njm_j}(\vec{r})\rangle$ corresponds to
$\kappa=\pm(j+1/2)$, $j=l\pm1/2$, and $m_j$ runs from $-j$ to $j$.
For $n=|\kappa|$, $\kappa$ only takes $j+1/2$, or $j=n-1/2$. ${\cal
N}=\int_0^{+\infty}r^2[f^2(r)+g^2(r)]dr$ is the normalization
constant. The exact solutions of $f(r)$ and $g(r)$ are
\cite{Bethe,AJP}
\begin{subequations}
 \label{fg}
 \begin{align}
f(r)&=\sqrt{Mc^2+E}[-\tilde{n}\;_1F_1(1-\tilde{n},2\nu+1,\rho)\nonumber\\
&+(Mc^2a\lambda+\kappa)_1F_1(-\tilde{n},2\nu+1,\rho)]\; \rho^{\nu-1}e^{-\rho/2},\\
g(r)&=\sqrt{Mc^2-E}[\tilde{n}\;_1F_1(1-\tilde{n},2\nu+1,\rho)\nonumber\\
&+(Mc^2a\lambda+\kappa)_1F_1(-\tilde{n},2\nu+1,\rho)]\;
\rho^{\nu-1}e^{-\rho/2},
 \end{align}
\end{subequations}
where $\tilde{n}=n-|\kappa|$, $\nu=\sqrt{\kappa^2-a^2}$,
$\rho=2r/\hbar c\lambda$,
$_1F_1(p;q;z)=1+\frac{p}{q}\frac{z}{1!}+\frac{p(p+1)}{q(q+1)}\frac{z^2}{2!}+\cdots$
is confluent hypergeometric function, $\lambda=1/\sqrt{M^2c^4-E^2}$,
and
\begin{subequations}
 \label{J2j}
 \begin{align}
\phi^A_{jm_j}&=\frac{1}{\sqrt{2l+1}}\left(\begin{array}{c}
\sqrt{l+m+1}\; Y_{lm}(\vartheta, \varphi)\\
\sqrt{l-m}\;Y_{l,m+1}(\vartheta,
\varphi)\end{array}\right),\\
\phi^B_{jm_j}&=\frac{1}{\sqrt{2l+3}}\left(\begin{array}{c}
-\sqrt{l-m+1}\;Y_{l+1,m}(\vartheta, \varphi)\\
\sqrt{l+m+2}\;Y_{l+1,m+1}(\vartheta, \varphi)\end{array}\right),
\end{align}
\end{subequations}
where $m=m_j-1/2$ and $Y_{lm}(\vartheta, \varphi)$ are the spherical
harmonics.

Let us now consider the following noncontextuality inequality:
\begin{equation}
\mathcal {I}=\langle AB\rangle+\langle BC\rangle+\langle
CD\rangle-\langle DA\rangle\leq2,\label{CHSH-like}
\end{equation}
where $A$, $B$, $C$, and $D$ are observables taking values $\pm1$,
and the pairs $(A,B)$, $(B,C)$, $(C,D)$, and $(D,A)$ contain
compatible observables. The inequality~(\ref{CHSH-like}) is similar
to the Clauser-Horne-Shimony-Holt (CHSH) Bell inequality, but does
not make a distinction between compatible observables which are
spacelike separated and those which are not \cite{GKCLKZGR10}. It is
therefore a noncontextuality inequality which can be tested on a
noncomposite system. Indeed, this inequality has been used for
testing contextuality on single photons~\cite{MWZ00} and single
neutrons~\cite{HLBBR03}.

The quantum contextuality of a relativistic hydrogen atom is stated
in the following theorem.


\emph{Theorem.} All eigenstates of $H_{\rm rha}$ violate the noncontextuality inequality (\ref{CHSH-like}).


\emph{Proof.} We introduce two sets of operators
\begin{subequations}
\begin{align}
 \vec{\Gamma}&=(\Gamma_x,\Gamma_y,\Gamma_z)=(\gamma^0,\gamma^2\gamma^0,i\gamma^2),\label{gam1}\\
 \vec{\Gamma}'&=(\Gamma'_x,\Gamma'_y,\Gamma'_z)=(\gamma^3\gamma^5,i\gamma^3\gamma^1,\gamma^5\gamma^1),\label{gam2}
\end{align}
\end{subequations}
where $\gamma^5=i\gamma^0\gamma^1\gamma^2\gamma^3$, and $\gamma$'s
are the Dirac gamma matrices in the Weyl basis:
\begin{subequations}
 \label{gamma}
 \begin{align}
&\gamma^0=\left(
\begin{matrix}
0& \openone\\
\openone &0\end{matrix}\right),\\
&\gamma^1=\left(
\begin{matrix}
0& \sigma_x\\
-\sigma_x&0\end{matrix}\right),\\
&\gamma^2=\left(
\begin{matrix}
0& \sigma_y\\
-\sigma_y&0\end{matrix}\right),\\
&\gamma^3=\left(
\begin{matrix}
0& \sigma_z\\
-\sigma_z&0\end{matrix}\right).
\end{align}
\end{subequations}
Due to the anticommutative relations
$\gamma^{i}\gamma^{j}=-\gamma^{j}\gamma^{i},\;i\neq j$, it is easy
to prove that $\vec{\Gamma}$ and $\vec{\Gamma}'$ commute.

The ground states are twofold degenerated. For the ground state
$|\psi^+_{1,\frac{1}{2},\frac{1}{2}}(\vec{r})\rangle$, we choose the
observables:
\begin{subequations}
\label{observ}
\begin{align}
 &A=\Gamma_x,\\
 &B=\frac{1}{\sqrt{2}}(\Gamma'_x-\Gamma'_z),\\
 &C=\Gamma_z,\\
 &D=-\frac{1}{\sqrt{2}}(\Gamma'_x+\Gamma'_z).
\end{align}
\end{subequations}
Quantum mechanically, the expectation value is given by
\begin{eqnarray}
\langle AB\rangle&=&\int_0^\infty r^2 dr \int_0^\pi\sin\vartheta
 d\vartheta\int_0^{2\pi} d\varphi \langle\psi|
AB|\psi\rangle,
\end{eqnarray}
with a similar expression for the other pairs. We obtain the quantum violation
\begin{equation}
\mathcal {I}^{\rm QM}=(1+\sqrt{1-a^2})\sqrt{2}\simeq2.82839,
\end{equation}
which is very close to $2\sqrt{2}$.

For the ground state
$|\psi^+_{1,\frac{1}{2},-\frac{1}{2}}(\vec{r})\rangle$, we choose
the observables $A=\Gamma_x$,
$B=\frac{1}{\sqrt{2}}(\Gamma'_z-\Gamma'_x)$, $C=\Gamma_z$, and
$D=\frac{1}{\sqrt{2}}(\Gamma'_x+\Gamma'_z)$, and obtain the same
value for the quantum violation $\mathcal {I}^{\rm QM}$. This proves
the case for the ground states.

For excited states with $\kappa>0$, we consider the following
observables: $A=\Gamma_y, B=-\sin\xi\;\Gamma'_y+\cos\xi\;\Gamma'_z,
C=\Gamma_z,D=\sin\xi\;\Gamma'_y+\cos\xi\;\Gamma'_z$.
Substituting them into Eq.~(\ref{CHSH-like}) and using
$\frac{1}{\mathcal {N}}\int_0^{\infty}r^2f^2(r) dr=(1+\mu)/2$,
$\frac{1}{\mathcal {N}}\int_0^{\infty}r^2g^2(r) dr=(1-\mu)/2$, and
$\mu=E/Mc^2$, the left-hand side of Eq.~(\ref{CHSH-like}) becomes
$2[-\frac{(2m+1)(\mu+2l+2)}{4l^2+8l+3}\cos\xi-\mu\sin\xi]$, which
can reach
\begin{subequations}
\begin{eqnarray}
\mathcal {I}^{\rm
QM}&=&2\sqrt{\mu^2+\frac{(2m+1)^2(\mu+2l+2)^2}{(4l^2+8l+3)^2}}\\
&>&
2\sqrt{\mu^2+\frac{(\mu+2l+2)^2}{(4l^2+8l+4)^2}}\label{b}\\
&>&2\sqrt{1+\frac{1-4a^2}{4(l+1)^2}}>2.\label{c}
\end{eqnarray}
\end{subequations}
In step (\ref{b}), we take $m=0$; in step (\ref{c}), $\mu$
takes the minimal value $\sqrt{1-a^2/\kappa^2}$ for $n=\kappa=l+1$.

For excited states with $\kappa<0$, we choose the same observables
as for the case of $\kappa>0$. Then, the left-hand side of Eq.~(\ref{CHSH-like}) becomes
$2[\frac{(2m+1)(-\mu+2l+2)}{4l^2+8l+3}\cos\xi-\mu\sin\xi]$, which
reaches the value
\begin{subequations}
\begin{eqnarray}
\mathcal {I}^{\rm
QM}&=&2\sqrt{\mu^2+\frac{(2m+1)^2(2l+2-\mu)^2}{(4l^2+8l+3)^2}}\\
&>&2\sqrt{\mu^2+\frac{(2l+1)^2}{(4l^2+8l+3)^2}}\label{bb2}\\
&=&2\sqrt{1+\frac{1}{(2l+3)^2}-\frac{a^2}{l^2+1}}>2.\label{bc2}
\end{eqnarray}
\end{subequations}
In step (\ref{bb2}), we take $m=0$; in step (\ref{bc2}),
$\mu^2$ takes the minimal value $\mu^2_{\rm
min}>1-\frac{a^2}{1+l^2}$ for $n=l+1$. Therefore, the
noncontextuality inequality (\ref{CHSH-like}) is always violated.
This completes the proof.\hfill \endproof

Let us point out that the above test of quantum contextuality is
state-dependent. By resorting to the Peres-Mermin
square~\cite{Mermin, Mermin-Peres}, we show that the quantum
contextuality for the relativistic hydrogen atom can also be
verified state-independently. The Peres-Mermin square contains nine
observables:
\begin{eqnarray}
P=\left(\begin{matrix} \Sigma'_z & \Sigma_z & \Sigma_z\Sigma'_z \\
\Sigma_x & \Sigma'_x & \Sigma_x\Sigma'_x\\
\Sigma'_z\Sigma_x & \Sigma'_x\Sigma_z & \Sigma_y\Sigma'_y
\end{matrix}\right),\label{PM}
\end{eqnarray}
where
$\vec{\Sigma}'=(\Sigma'_x,\Sigma'_y,\Sigma'_z)=\vec{\sigma}\otimes\openone$.
Note that observables in the same row or column mutually commute.
They violate the following noncontextuality
inequality~\cite{Cabello08}:
\begin{eqnarray}
\langle P_{11}P_{12}P_{13}\rangle+\langle
P_{21}P_{22}P_{23}\rangle+\langle P_{31}P_{32}P_{33}\rangle+\nonumber\\
\langle P_{11}P_{21}P_{31}\rangle+\langle
P_{12}P_{22}P_{32}\rangle-\langle P_{13}P_{23}P_{33}\rangle\leq 4,
\label{PM1}
\end{eqnarray}
where $P_{ij}\;(i,j=1,2,3)$ are the corresponding matrix entries in
Peres-Mermin square, and are dichotomic observables which commute
with one another in the same correlator. One can verify that, for
noncontextual theories, the upper bound of the inequality is 4.
However, quantum mechanics gives 6, regardless of details of the
states. This state-independent advantage readily allows one to
verify quantum contextuality for arbitrary four-spinor states
(\ref{E1}a) and (\ref{E1}b).


\emph{Remark 1.} In nonrelativistic quantum mechanics, there are
no enough compatible observables for a single spin-$1/2$ particle to
establish the inequalities (\ref{CHSH-like}) and (\ref{PM1}). Here
we provide an intuitive reason why it is possible for the case in
relativistic quantum mechanics. Let us focus on the operator
$\vec{\Sigma}=\openone\otimes\vec{\sigma}$. One finds that it
possesses a very nice property: the eigenvalues are $+1$ and $-1$
(or equivalently its square is a $4\times 4$ unit matrix). Moreover,
one easily observes that any operator of the form
$\mathcal{O}\otimes \openone$ commutes with $\vec{\Sigma}$. If one
requires the eigenvalues of $\mathcal{O}\otimes \openone$ are also
$+1$ and $-1$, then the general form of the operator is
\begin{equation}
 \label{HRHA5}
\vec{\Sigma}'=\mathcal{V}\vec{\sigma}\mathcal{V}^\dag\otimes
\openone,
\end{equation}
with $\mathcal{V}$ the $2\times 2$ unitary matrix. The components of
$\vec{\Sigma}$ and $\vec{\Sigma}'$ can be used to construct the nine
observables in the Mermin-Peres square [see Eq. (\ref{PM}), where we
have simply set $\mathcal{V}=\openone$], therefore the standard KS
theorem is applicable for the relativistic spin-1/2 particle by
violation of the state-independent noncontextuality inequality
(\ref{PM1}).

Moreover, it can be verified directly that the operator
$\vec{\Sigma}\cdot \vec{n}$ commutes with the operator
 $\vec{\Sigma}'\cdot\vec{n}'$, where $\vec{n}$, $\vec{n}'$ are some directions in the three-dimensional space. In general, up to a unitary transformation $\mathcal{U}$,
 the two operators
 \begin{subequations}
\begin{align}
\vec{\Gamma} \cdot \vec{n} =\mathcal{U} \; \vec{\Sigma} \; \mathcal{U}^\dag\cdot \vec{n},\\
\vec{\Gamma}' \cdot \vec{n}' = \mathcal{U} \; \vec{\Sigma}' \;
\mathcal{U}^\dag\cdot \vec{n}',
\end{align}
\end{subequations}
are commutative. Thus $(\vec{\Gamma} \cdot \vec{n}, \vec{\Gamma}'
\cdot \vec{n}')$ is a compatible pair of observables. By choosing an
appropriate unitary transformation $\mathcal{U}$, one may arrive at
the operators $\vec{\Gamma}$ and $\vec{\Gamma}'$ as in Eq.~(\ref{gam1}) and Eq. (\ref{gam2}). Then the construction of
observables $A, B, C, D$ in the inequality (\ref{CHSH-like}) is as
follows:
\begin{subequations}
\begin{align}
A= \vec{\Gamma} \cdot \vec{n}_a, \;\;\;\; C= \vec{\Gamma} \cdot \vec{n}_c, \\
B= \vec{\Gamma}' \cdot \vec{n}'_b, \;\;\;\; D= \vec{\Gamma}' \cdot
\vec{n}'_d. \end{align}
\end{subequations}
It is easy to check that eigenvalues of $A, B, C, D$ are $\pm 1$,
and $(A,B)$, $(B,C)$, $(C,D)$, $(D,A)$ are compatible pairs. For the
observables (\ref{observ}a)--(\ref{observ}d), we have chosen the
directions as $\vec{n}_a=(1, 0, 0)$, $\vec{n}'_b=(\cos\theta, 0,
-\sin\theta)$, $\vec{n}_c=(0, 0, 1)$, $\vec{n}'_d=(-\cos\theta, 0,
-\sin\theta)$. Thus the standard KS theorem is also applicable for
the relativistic spin-1/2 particle by violation of the CHSH-like
noncontextuality inequality of Eq.~(\ref{CHSH-like}).

\emph{Remark 2.} The eigenstates of the observables
(\ref{observ}a)--(\ref{observ}d) are superpositions of eigenstates of
$H_{\rm rha}$, whose eigenenergies (\ref{E}) are all positive. This
makes the observables (\ref{observ}a)--(\ref{observ}d) in
principle measurable. Let us take observable $A$ for an example.
Assume its eigenstates are $|u_i\rangle, (i=1,2,3,4)$, then each
$|u_i\rangle$ can be expanded as $|u_i\rangle=\sum_\ell (c^{+}_\ell
|\psi^{+}_\ell\rangle+c^{-}_\ell |\psi^{-}_\ell\rangle)$, with
$|\psi^{\pm}_\ell\rangle$ eigenstates (\ref{E1}a) and
(\ref{E1}b), respectively, and $\ell$ denoting indices $njm_j$.


\section{Quantum contextuality for the free Dirac
electron and the relativistic Dirac oscillator}


Let us discuss here the KS contextuality for a free Dirac
electron. For $V(r)=0$, we have the Hamiltonian of
the free Dirac electron from Eq.~(\ref{HRHA1}) as
\begin{equation}
 \label{HRHA2}
 H_{\rm e}=c\ \vec{\alpha}\cdot \vec{p}+\beta Mc^2.
\end{equation}
For simplicity, we assume that the electron is moving in the
$z$ direction. For a given momentum $\vec{p}=\hbar k\hat{e}_z$,
energy $E=\sqrt{M^2c^4+\hbar^2c^2k^2}$, and helicity
$\vec{\Sigma}\cdot\vec{p}=\pm\hbar k$, the four-spinor eigenstate
reads
\begin{eqnarray}
|\Psi^\pm_e(k)\rangle=\frac{1}{\sqrt{\mathcal {N}_e}}\left(
\begin{matrix}
\chi^{\pm}\\ \frac{c\hbar k}{Mc^2+E}\chi^{\pm}
\end{matrix}\right)e^{ikz},\label{free-elec}
\end{eqnarray}
where the two-spinors $\chi^{+}=\left(\begin{matrix} 1\\
0
\end{matrix}\right),\chi^{-}=\left(\begin{matrix} 0\\
1
\end{matrix}\right)$ are the spin-up and spin-down states of $\sigma_z$, respectively, and $\mathcal {N}_e=2E/(Mc^2+E)$ is the normalization constant.
If one adopts the following observables:
\begin{subequations}
 \label{obse-electron}
\begin{align}
 &A'=\gamma^0,\\
 &B'=(\cos\theta\gamma^3+\sin\theta\gamma^1)\gamma^5,\\
 &C'=i\gamma^2,\\
 &D'=(-\cos\theta\gamma^3+\sin\theta\gamma^1)\gamma^5,
\end{align}
\end{subequations}
with $\theta=\arctan\left(\frac{Mc^2}{E}\right)$. Then for the state
$|\Psi^+_e(k)\rangle$, the quantum prediction reads
\begin{eqnarray}
&& \mathcal {I}^{\rm
QM}=2\sqrt{2-\frac{c^2\hbar^2k^2}{E^2}}=2\sqrt{2-\frac{v^2}{c^2}},
\label{qv-e}
\end{eqnarray}
which beats the upper bound of the noncontextuality inequality
(\ref{CHSH-like}) for any $v< c$, where $v=c^2|\vec{p}|/E$ is the
velocity of the electron. However, this does not imply the KS
contextuality of $H_{\rm e}$ is identified. Because the eigenstates
of observables (\ref{obse-electron}a)--(\ref{obse-electron}d) are
superpositions of both positive and negative energy wave functions
(i.e., both electron and positron states) for nonzero momentum, this
hinders the measurability of the observables~\cite{wick,sup}. The same
discussion also applies for the relativistic Dirac oscillator,
\begin{equation}
 \label{HRHA3}
 H_{rdo}=c\ \vec{\alpha}\cdot (\vec{p}-iM\omega\beta\vec{r})+\beta Mc^2,
\end{equation}
whose eigenenergies can also be positive and
negative~\cite{DiracOsc}.


\section{Conclusions}


Although the existence of KS contextuality for a single spin-1/2
particle remains a disputed problem within nonrelativistic quantum
mechanics, here we have shown that KS contextuality for an electron
moving in the Coulomb potential naturally emerges within a
relativistic treatment. Within this approach, we have explored the
quantum contextuality of the relativistic hydrogen atom through
violations of noncontextuality inequalities. We have proven that all
eigenstates of the atom violate noncontextuality inequalities. This
confirms that contextuality exists in the domain of relativistic
quantum mechanics.

A distinction between relativistic and nonrelativistic quantum
mechanics is that negative energies of antiparticles may emerge in
the former. Given a relativistic Hamiltonian, if a Hermitian
operator cannot be expanded by the eigenstates of Hamiltonian with
positive energy alone, then it cannot be viewed as a measurable
observable. In nonrelativistic quantum mechanics, the KS
contextuality only depends on the dimension of the system, i.e., any
system with $d\geq 3$ has contextuality. Nonetheless, in the
framework of relativistic quantum mechanics, when investigating the
KS contextuality one has to additionally take the measurability of
observables into account. Although to some extent the four-spinor
states can be viewed as a four-dimensional system, the dimension of
the relativistic system itself does not guarantee the existence of
the KS contextuality. We expect further developments of
contextuality in relativistic quantum mechanics in the near future.


\begin{acknowledgments}
We thank V.\ Vedral for valuable discussion. J.\ L.\ C. is supported by
National Basic Research Program (973 Program) of China under Grant
No.\ 2012CB921900 and the NSF of China (Grant Nos.\ 10975075 and
11175089). A.\ C. is supported by the Spanish Ministry of Economy 
and Competitiveness Project No.\ FIS2011-29400 and the Wenner-Gren 
Foundation. This work is also partly supported by the National 
Research Foundation and the Ministry of Education of Singapore 
(Grant No.\ WBS:
R-710-000-008-271).
\end{acknowledgments}




\begin{thebibliography}{99}
\bibitem{sup}
Let us take observable $C'$ for an example. Its eigenstates are
$|u_1\rangle=(-1/\sqrt{2},0,0,1/\sqrt{2})^T$,
$|u_2\rangle=(0,1/\sqrt{2},1/\sqrt{2},0)^T$,
$|u_3\rangle=(1/\sqrt{2},0,0,1/\sqrt{2})^T$,
$|u_4\rangle=(0,-1/\sqrt{2},1/\sqrt{2},0)^T$. For nonzero momentum,
each $|u_i\rangle$ cannot be expanded only by
$|\Psi^{+}_e(k)\rangle$ and $|\Psi^{-}_e(k)\rangle$, whose
corresponding energies are positive.
\end{thebibliography}
\end{document}